\documentclass[prl,preprint]{revtex4-1}
\usepackage{graphicx}
\usepackage{dcolumn}
\usepackage{bm}
\usepackage{amsmath}

\begin{document}

\title{Photo-spin voltaic effect and photo-magnetoresistance in proximized platinum}
\author{D. Li}
\author{A. Ruotolo}
\email{aruotolo@cityu.edu.hk}
\affiliation{Department of Physics and Materials Science, Centre for Functional Photonics, City University of Hong Kong, Kowloon, Hong Kong SAR, China}

\date{\today}

\begin{abstract}
Spin orbit coupling in heavy metals allows conversion of unpolarized light into an open-circuit voltage. We experimentally prove that this photo-spin voltaic effect is due to photo-excitation of  carriers in the proximized layer and can exist for light in the visible range. While carrying out the experiment, we discovered that, in closed-circuit conditions, the  anisotropic magnetoresistance of the proximized metal is a function of the light intensity. We name this effect photo-magnetoresistance. A magneto-transport model is presented that describes the change of magnetoresistance as a function of the light intensity.  
\end{abstract}

\maketitle 

Photovoltaic conversion is the result of two concurrent physical processes: photo-generation and separation of carriers.  In semiconductors, generation relies on the absorption of a photon with energy larger than the band-gap to transfer an electron from the valence band to the conduction band or a hole from the conduction band to the valence band. Separation of charges with different sign relies on the built-in electric field at a $\it{pn}$ interface junction. 

In spintronics, carriers are discriminated based on their spin angular momentum, rather than their charge. While separation between electrons with spin-up and spin-down can readily be achieved by using inverse spin-Hall effect \cite{Saitoh:APL:2006}, photo-generation of electrons with different populations of spins remains challenging. The spin-galvanic effect can be used to generate an imbalance of spins via absorption of circularly polarized light \cite{Ganichev:Nat:2002}. Yet, it still makes use of semiconductors and the generated voltage is usually so small to require lock-in techniques to be detected \cite{Ando:APL:2010}. The recently discovered photo-spin-voltaic effect is unique in that it uses unpolarized light to obtain photo-voltaic conversion in metals \cite{Ellsworth:NatPhys:2016}. It relies on the excitation of polarized carriers in the proximized layer of a metal with large spin orbit coupling (SOC) in contact with a ferromagnetic insulator, with the separation taking place because of inverse spin-Hall effect. In this pioneering experiment conversion was achieved by using infrared light. Moreover, the effect was interpreted based on density functional theory calculations for optical absorption and spin diffusion analysis but no direct experimental evidence was given to support this interpretation.
  
We here show that photo-spin voltaic effect can exist in the visible range. Moreover, we provide a direct experimental evidence of the physical origin of the effect by measuring the  anisotropic magnetoresistance (MR) in the same configuration under different light intensities. Our experiment supports the original interpretation of the effect. From a different prospective, we demonstrate that MR in a proximized metal is light-dependent. An analytical model is presented that describes this effect.

\begin{figure}
\includegraphics[width=12cm]{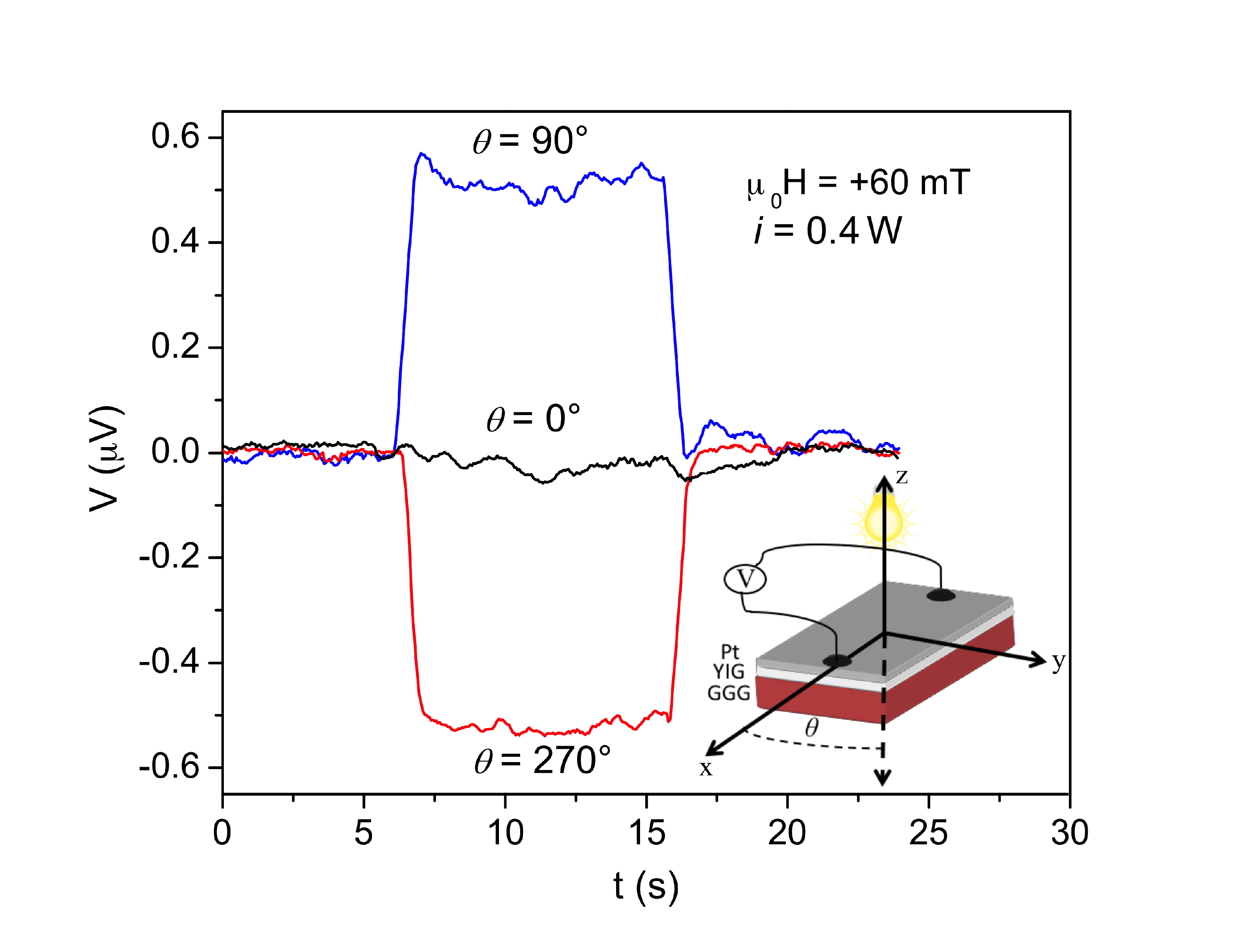}
\caption{\label{PSV} Photo-spin voltaic effect under visible light illumination. The inset shows the measurement configuration.} 
\end{figure}

The system we studied was an yttrium ion garnet Y$_{3}$Fe$_{5}$O$_{12}$/platinum (YIG/Pt) bilayer. A 5-$\mu$m thick, single-crystal $<111>$YIG, grown by liquid phase epitaxy on $<111>$ gadolinium gallium garnet (GGG) wafer, was purchased. The wafer was diced in samples of size 1 cm $\times$ 0.5 cm. In order to avoid any possible magnetic contamination or diffusion of Pt into the YIG, pulsed laser deposition at room temperature was used to deposit 3 nm of Pt. The resistivity of the Pt layer was measured to be $\rho_{\text{Pt}} = 2.1 \times 10^{-6}~\Omega$m, which is the expected resistivity for a ultra-thin Pt film of the same thickness \cite{Avrekh:ASS:2000}. The dye was wire-bonded to a chip holder and placed in between the poles of an electromagnet. The schematic of the experimental setup is shown in Fig.~\ref{PSV}.
The magnetic field is always applied along a fixed $y$-direction and the angle of the light illumination is set along the $z$-direction. The sample holder can be rotated in the $x-y$ plane in such a way that the voltage signal is measured with a nano-voltmeter between the two bonded contacts at an angle $\theta$. The nanovoltmeter was set to operate with a low-pass filter with frequency cut-off $f_C = 18$ Hz to reduce the noise to less than $\pm~0.05~\mu$V.

At first, an infrared laser with diffuser (beam diameter of 1 cm$^2$), as well as an halogen lamp and a long-pass infrared filter, were used to illuminate the sample. The system did not respond to the optical excitation. Rather, infrared light resulted in an increase of the sample temperature and transitory appearance of slow thermal effects associate with a temperature gradient, such as the spin-Seebeck effect and Nernst effects \cite{Uchida:Nature:2008,Uchida_JPCM_2014,Huang:PRL:2011,Avery:PRL:2012}. All the measurements shown in the following were taken by using an optical fiber illuminator with an as-measured wavelength spectrum covering the entire visible range from $\lambda$ = 400 nm (violet) to $\lambda$ = 700 nm (red) (see supplementary materials). In the following, the light intensity is the power at the sample obtained by removing all the filters from the sensor of the power meter and multiplying by the ratio between the sample area and sensor area. We could reproduce all the results reported in Ref. \cite{Ellsworth:NatPhys:2016}, but in our case the system responded to excitation in the visible, rather than infrared, range. In brief, if the YIG was magnetized along the $y$-axis, an open-circuit voltage of the order of $\sim \mu$V was detected along the $x$-axis (see Fig.~\ref{PSV}) when light was shed along the $z$-axis. The change of voltage was far too sharp for the effect to be thermal. Besides, the voltage magnitude did not change with the temperature, for temperatures as low as 10 K (see supplementary materials). The voltage magnitude was found to be a linear function of the light intensity and to flip its sign if the magnetization was reversed. No signal could be detected if the magnetization was aligned along the $x$-axis. This means that for an arbitrary angle $\theta$, the $x$-component of the magnetization vector will not contribute. As a consequence,  the voltage magnitude changes with the sine of the angle $\theta$, because so does the component of the magnetization along the $y$-axis.

This behavior is a clear signature that inverse spin-Hall effect is the electro-motive force that separates the spin-polarized charges \cite{Hoffman:IEEE:2013}. The open question is why an imbalance of spin exists in the first place, since Pt is a paramagnet and YIG is an insulator. Since the same behavior is observed when YIG is replaced by other magnetic insulators \cite{Ellsworth:NatPhys:2016,Li:APL:2014}, one can speculate that it must come from the ferromagnetically proximized layer in the Pt. Providing an experimental evidence to this assumption was the original motivation of our work.

If the Pt is thinner than the light penetration depth, then spin-polarized carriers in the proximized layer can be excited to produce a pure spin-current along the $z$-axis \cite{Li:APL:2014}. Light penetration depth is wavelength-dependent: $\delta = \sqrt{\rho\lambda / \pi \mu c}$, where $\rho$ is the resisitivity of the metal, $c$ is the speed of light in vacuum and $\mu$ is the magnetic permeability. In our case,  $\delta$ = 2.6 nm (resp. 3.5 nm) for $\lambda = 400$ nm (resp. 700 nm). Therefore visible light can reach the proximized layer if the thickness of the Pt film is 3 nm. If we compare our results with those in Ref. \cite{Ellsworth:NatPhys:2016}, we conclude that the experiment must be extremely sensitive to thickness and resistivity of Pt. A systematic study of the effect as a function of the Pt thickness requires an accurate control of the film thickness. Besides, for different thicknesses, the system would respond at different wavelengths, and a wavelength-dependent study is at the moment not at reach, given the weak magnitude of the voltage elicited. We can point out here that there is no physical reason why the effect should only exist for infra-red light. Rather, provided that the light can reach the interface, lower wavelengths correspond to higher photon energies and, therefore, higher photo-excitation efficiency.

\begin{figure}
\includegraphics[width=12cm]{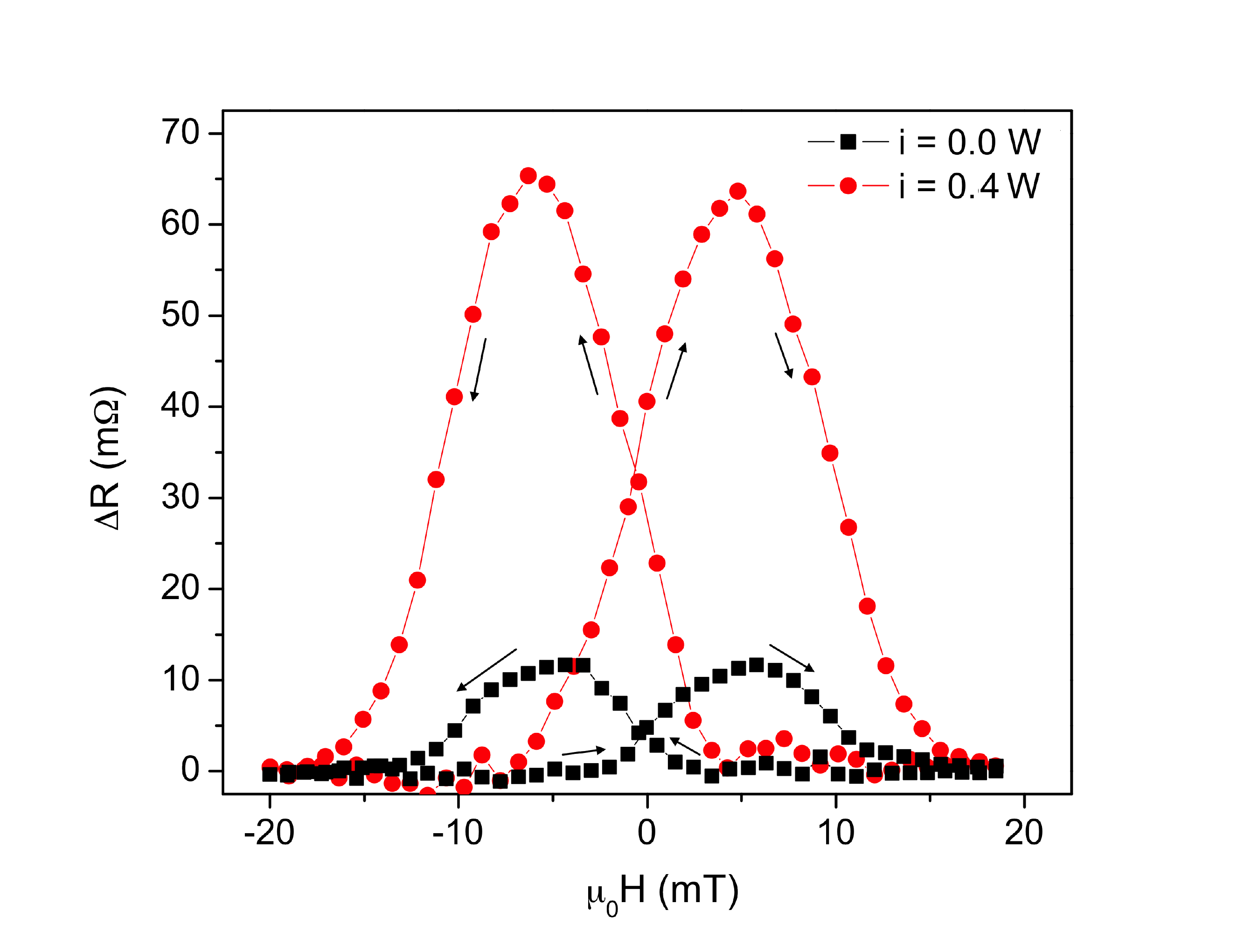}
\caption{\label{PMR} Magnetoresistance for different light illumination.} 
\end{figure}

In the proximized layer, an exchange interaction exists between conduction electrons that decays with the distance from the interface. If the light changes spin-population distribution in open-circuit conditions, it should also change the magneto-transport characteristics of the Pt in closed-circuit conditions. Proximized Pt is known to show MR \cite{Huang:PRL:2012,Lu:PRL:2013,Nakayama:PRL:2013}. Therefore we measured the MR of the films under different light illumination. We indeed found the anisotropic MR to be light-dependent. Fig.~\ref{PMR} shows the MR of the sample in dark and bright condition with light intensity $i = 0.4$ W. 
The sample geometry is the same as that used in open-circuit conditions but a current was injected along the $x$-axis. The magnetic field was swept along the $y$-axis. The magnetoresistance ratio was found to be independent on the injected current. A maximum current of $200~\mu$A was injected. This, together with the large size of the sample, excludes contribution from spin-Hall megnetoresistance (SMR) \cite{Nakayama:PRL:2013}. For the following, it is important to understand that the total resistance of the samples was $R \sim 2$~k$\Omega$, therefore the $\Delta R$ in Fig.~\ref{PMR} corresponds to a magnetoresistance $\Delta R/R$ of less than 0.01\%, far smaller than the anisotropic MR of a ferromagnetic metal. This means that the average magnetic moment induced in the Pt by the proximity with the YIG is much smaller than the magnetic moment of common ferromagnetic metals. By using x-ray magnetic circular dichroism, an average induced magnetic moment of 0.05 $\mu_\text{B}$ (with $\mu_\text{B}$ Bohr magneton) at room temperature has been estimated in a Pt (1.5 nm) film proximized by YIG \cite{Lu:PRL:2013}.

\begin{figure}
\includegraphics[width=12cm]{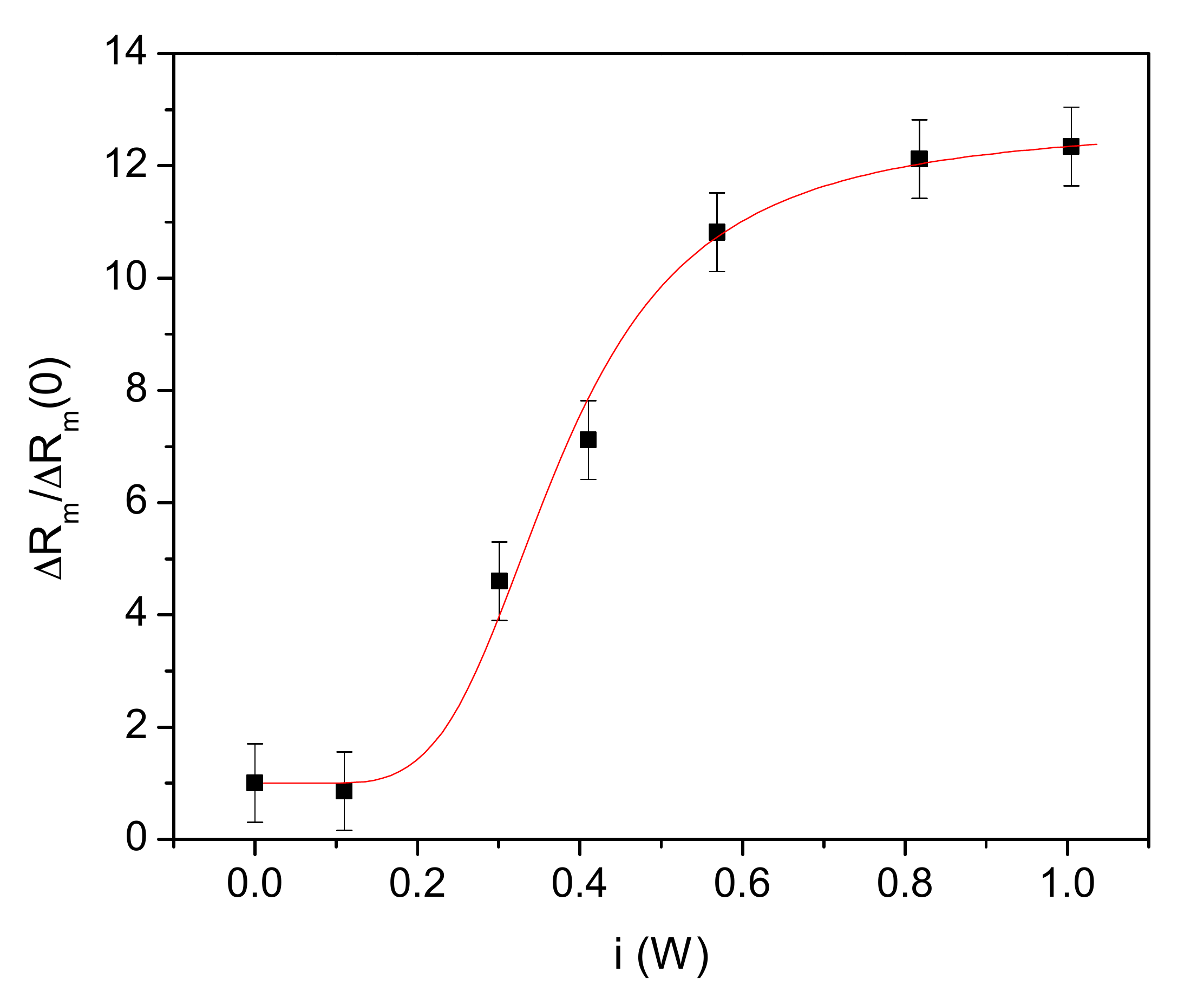}
\caption{\label{dRvsI} Maximum change of resistance as a function of light intensity. The line is the best fit with Eq.~\ref{expected}} 
\end{figure}

In Fig.~\ref{dRvsI} we plot the maximum change of resistance ($\Delta R_{m}$) as a function of the light intensity. $\Delta R_{m}$ was found to increase with light intensity. In particular, for low values of the light intensity, $\Delta R_{m}$ does not change significantly. A significant increase exists for intermediate intensities, after which $\Delta R_{m}$ approaches a limit value. The change of resistance cannot be simply ascribed to the additional photo-induced voltage because the latter is linearly proportional to the light intensity.

\begin{figure}
\includegraphics[width=12cm]{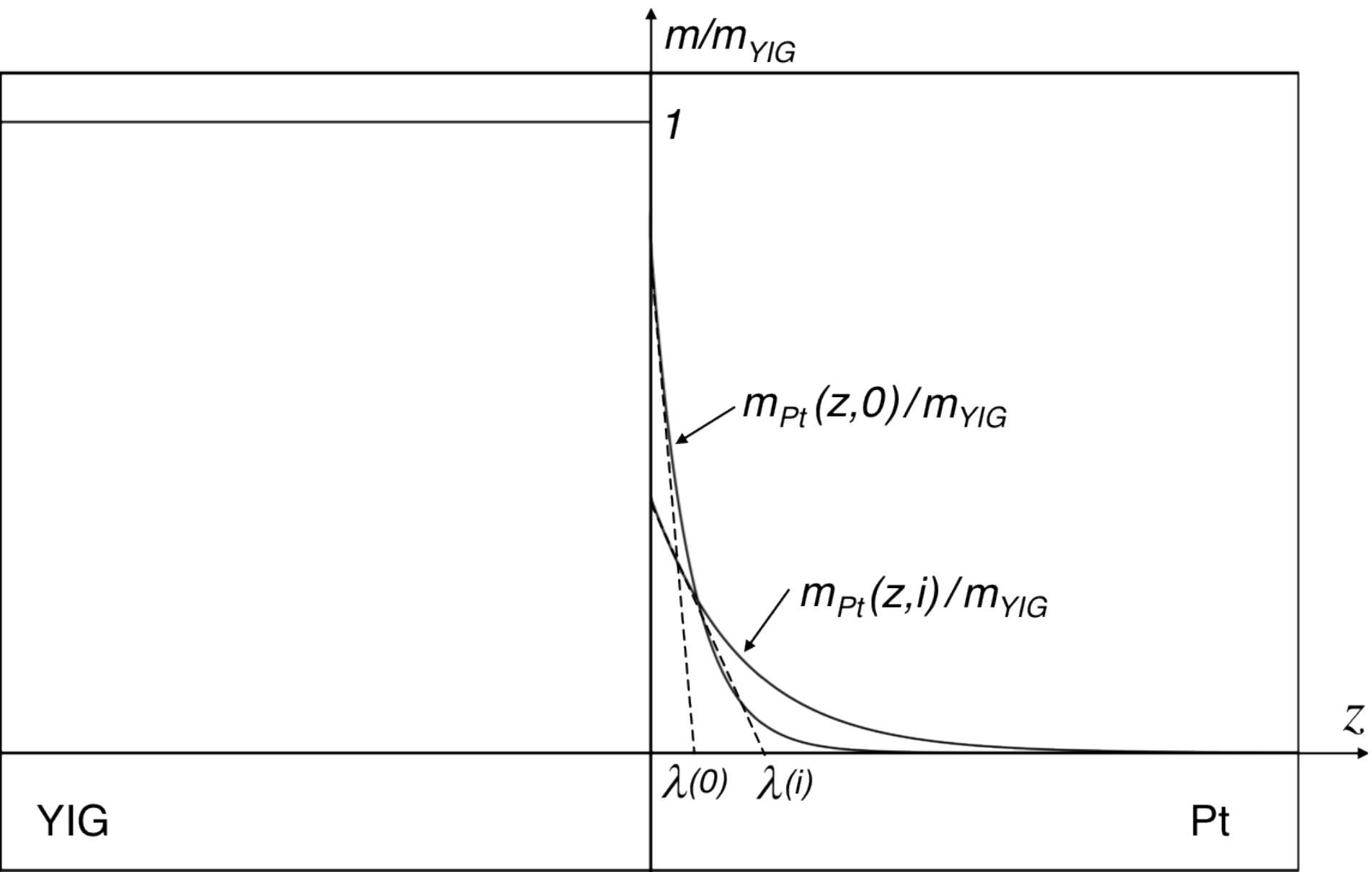}
\caption{\label{mvsz} Schematic representation of magnetic proximity effect in YIG/Pt biayers.} 
\end{figure}

In the following, a magneto-transport model is presented that well describes the effect. The model is based on the assumption that Pt is proximized in a similar fashion as a normal metal would be in contact with a superconductor \cite{DeGennes:RMP:1964}, with the order parameter being, in this case, the magnetic order (see Fig. \ref{mvsz}). 
The order parameter is in principle suppressed in the ferromagnet over a certain coherence length near the interface. Yet, YIG is known not to experience a significant reduction of the magnetic moment near the surface (dead layer effect \cite{Ruotolo:APL:2006}), or interface with metals \cite{Niyaifar:JMMM:2016}. Therefore, a constant magnetic moment $m_{YIG}$ can be assumed in the ferromagnet. Instead, the order parameter decays exponentially in the normal metal. The decay constant is the coherence length $\xi$. Since the anisotropic magnetoresistance is directly dependant on the induced magnetic moment, light must change the magnetic profile, i.e. $m$ and $\xi$ must be a function of $i$. 
Under these assumption, the induced magnetic moment in the Pt can be written as:

\begin{equation}
m(z,i) = m(0,i)e^{-\frac{z}{\xi{(i)}}} = m_0(i)e^{-\frac{z}{\xi{(i)}}}
\label{moment}
\end{equation}
where $z$ is the distance from the interface ($z > 0$) and $i$ is the light intensity.

Unpolarized light cannot polarize charges, which can be expressed as:
\begin{equation}
\int_0^t m(z,i) dz = k~~~~~\forall i
\label{ansatz}
\end{equation}
where the constant $k$ is the area under the curve in Fig. \ref{mvsz}. 

By replacing eq.~\ref{moment} in eq.~\ref{ansatz}, one finds:

\begin{equation}
\int_0^t m(z,i) dz = m_0(i)\xi(i)\left[1-e^{-\frac{t}{\xi(i)}}\right] = k~~~~~\forall i
\label{cond}
\end{equation}

Let us notice that $m(0,i) = m_0(i)$ depends also on the light penetration depth, and therefore on film thickness and wavelength. Yet, in our case, the thickness and the light spectrum are kept constant throughout the experiment. 

The resistivity of the proximized Pt can be written as \cite{Fert:JMMM:1999}:
\begin{equation}
\rho(z,i) = \rho_0 + \Delta\rho_m (z,i) \cos^2\theta
\end{equation}
where $\theta$ is the angle between the directions of the current and magnetization. The maximum change of resistivity $\Delta\rho_m$ depends on the magnetic order, \emph{i.e.} it is a monotonic function of the magnetic moment, $\Delta\rho_m = g(m)$, and satisfies the conditions: $\delta g / \delta m > 0$ (it increases with magnetic moment) \& $g(m = 0) = 0$  (non-magnetic metals do not show anisotropic magnetoresistance). If the induced average magnetic moment $<m(z,i)>$ is small as compared to that of a ferromagnetic metal, which is certainly the case for Pt, one can assume, as a first approximation, a liner relationship between $\Delta\rho_m$ and $m$, $\Delta\rho_m = g(m) \approx \alpha m$ with $\alpha$ the constant of proportionality:

\begin{equation}
\Delta\rho_m (z,i) = \alpha m(z,i) = \alpha m_0(i)e^{-\frac{z}{\xi(i)}}
\end{equation}

A layer of length $l$, width $w$ and infinitesimal thickness $dz$ will offer a maximum change of resistance $d\Delta R_m$:

\begin{equation}
d \Delta R_m (z,i) = \Delta \rho_m (z,i) \frac {l}{w} \frac {1}{dz} = \alpha m_0(i)  e^{-\frac{z}{\xi(i)}} \frac {l}{w} \frac {1}{dz}
\end{equation}

The current is conducted in parallel in the Pt layers of infinitesimal thickness $dz$, therefore the total change of conductance $\Delta G_m (i)$ must be the sum of the changes of conductance in the layers of infinitesimal thickness $dz$:

\begin{equation}
d \Delta G_m (z,i) = \frac{1}{d \Delta R_m (z,i)} = \frac{e^{\frac{z}{\xi(i)}} }{\alpha m_0(i)}   \frac {w}{l} dz
\end{equation}

\begin{equation}
\Delta G_m (i) = \int_{0}^{t} \frac{e^{\frac{z}{\xi(i)}}}{\alpha m_0(i)}  \frac {w}{l} dz = \frac{\xi(i)}{\alpha m_0(i)}  \frac {w}{l}\left[e^{\frac{t}{\xi(i)}}-1\right]
\end{equation}

The maximum change of resistance as a function of the light intensity is therefore:

\begin{equation}
\Delta R_m (i) =  \frac{\alpha m_0(i)}{\xi(i)}  \frac {l}{w}\left[\frac{1}{e^{\frac{t}{\xi(i)}}-1}\right]
\end{equation}

By normalizing to the maximum change of resistance in the dark:

\begin{equation}
\frac{\Delta R_m (i)}{\Delta R_m (0)} =  \frac{\alpha m_0(i)}{\xi(i)}  \frac{\xi(0)}{\alpha m_0(0)}\left[\frac{e^{\frac{t}{\xi(0)}}-1}{e^{\frac{t}{\xi(i)}}-1}\right]
\end{equation}

and using the condition in eq. \ref{cond}:

\begin{equation}
\frac{\Delta R_m (i)}{\Delta R_m (0)} =   \frac{\xi^2(0)}{\xi^2(i)} \frac{\cosh (t/\xi(0))-1}{\cosh (t/\xi(i))-1} 
\label{expected}
\end{equation} 

\begin{figure}
\includegraphics[width=12cm]{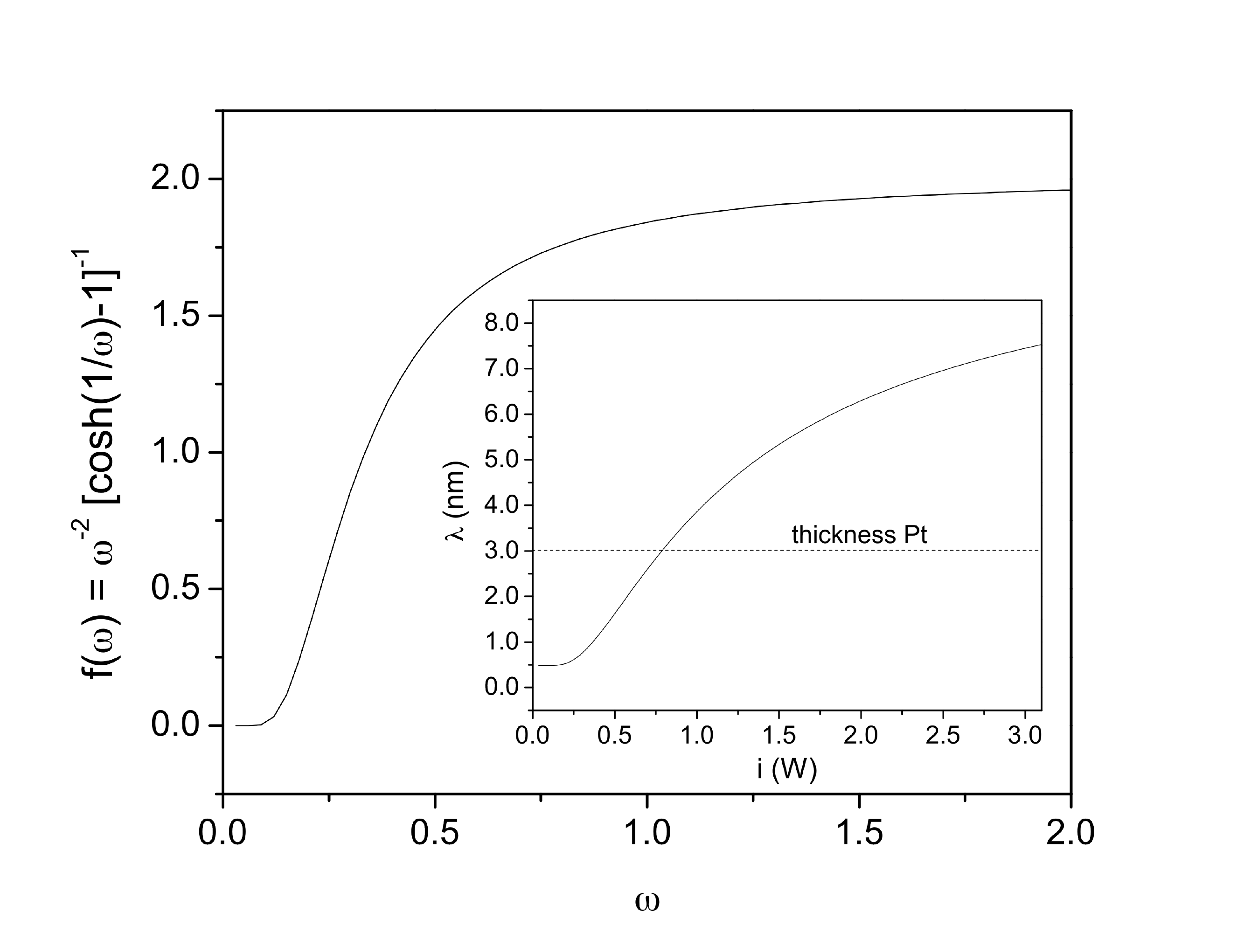}
\caption{\label{fx} Calculated trend for magnetoresistance vs light intensity. Inset shows the spin depth as a function of light intensity that gives the best fit of the experimental data.} 
\end{figure}

Indeed, the function $y (\omega) = (1/\omega^2) (1/(\cosh (1/\omega) -1))$, plotted in Fig.~\ref{fx}, is qualitatively in good agreement with the trend in our experimental data in Fig.~\ref{dRvsI}. A quantitative fit required an heuristic determination of the function $\xi = \xi(i)$. Polynomial and logarithmic functions resulted in a poor fit. The experimental data were instead well-fitted by assuming $\xi (i) = \xi(0) + \kappa e^{-\iota/i}$. The best fit in Fig.~\ref{dRvsI}, was achieved with $\xi(0) = 0.49$ nm, $\kappa = 10.0$ nm and $\iota = 1.08$ W. The inset in Fig.~\ref{fx} shows the function $\xi = \xi(i)$ for the same values of the parameters. Assuming an average moment in dark conditions $<m(z,0)> = 0.05~\mu_{\text{B}}$ \cite{Lu:PRL:2013} and using eq.~\ref{cond}, we estimated a moment at the interface $m_0(0) = 0.15~\mu_{\text{B}}$. 

For the maximum value of the intensity in our experiment, $\xi$ reaches the thickness of the film, after which our model will no longer be valid. Clearly the simple model we have presented here is very limited in the conditions and could not be applicable for ultra-thin films (thickness $< \approx$ 1 nm). Yet, it is very effective in fitting our data because, according to eq.~\ref{cond}, with $\xi$ increasing with intensity, the magnetic moment decreases much faster. The model is meant to inspire further theoretical studies on the ferromagnetic proximity effect.

In conclusion, we were able to experimentally prove that photo-spin voltaic effect is due to excitation of polarized carriers in the proximized layer of the metallic film. We also proved that the effect is observable when the light source is in the visible spectrum. In closed-circuit conditions, the proximized metal offers an anisotropic magnetoresistance that is light dependent. This photo-magnetoresistive effect could be well described by a simple magneto-transport model. By fitting the experimental data with the derived model we were able to estimate the the thickness of the proximized layer. An increase of the moment induced in the proximized layer, for instance by using $<100>$YIG \cite{Liang:ACS:2016}, and the use of materials with stronger spin-orbit coupling, such as topological insulators, could lead to optically-tunable magnetic sensors.

\section{Supplementary Material}
See supplementary materials for the measured spectrum of the light source used in the experiment and for the measurements of spin-voltaic effect at low temperature.

\section{Acknowledgment}
This study was funded by the Science, Technology and Innovation Commission of Shenzhen Municipality (Project no. JCYJ20170307091130687) and by the City University of Hong Kong (Project no.  CityU 7004898).

\newpage
\clearpage

\newpage
\clearpage

\newpage
\clearpage

\newpage
\clearpage

\newpage

\end{document}